\documentclass[runningheads]{cl2emult}

\usepackage{makeidx}  
\usepackage{graphicx} 
\usepackage{subeqnar} 
\usepackage{multicol} 
\usepackage{cropmark} 
\usepackage{eso}      
\makeindex            

\def\ltsima{$\; \buildrel < \over \sim \;$}
\def\lsim{\lower.5ex\hbox{\ltsima}}
\def\gtsima{$\; \buildrel > \over \sim \;$}
\def\gsim{\lower.5ex\hbox{\gtsima}}
\begin{document}
\title*{The BMW Deep X--ray Cluster Survey}
\toctitle{The BMW Deep X--ray Cluster Survey}
%
%
\titlerunning{The BMW Deep X--ray Cluster Survey}
%
\author{Luigi Guzzo\inst{1}
\and Alberto Moretti\inst{1}
\and Sergio Campana\inst{1}
\and Stefano Covino\inst{1}
\and Ian Dell'Antonio\inst{2}
\and Davide Lazzati\inst{1}
\and Marcella Longhetti\inst{1}
\and Emilio Molinari\inst{1}
\and Maria Rosa Panzera\inst{1}
\and Gianpiero Tagliaferri\inst{1}
}
\authorrunning{Luigi Guzzo et al.}
%
%
\institute{Osservatorio Astronomico di Brera, Via Bianchi 46, 
I-23807 Merate, Italy 
\and Physics Department, Brown University, Box 1843,
Providence, RI, USA}

\maketitle              

\begin{abstract}
We briefly describe the main features and first results of the BMW survey
of serendipitous X--ray clusters, based on the still unexploited
ROSAT--HRI archival observations.  The sky coverage, surface density
and first deep CCD images of the candidates indicate that this sample
can represent an excellent complement to the existing PSPC deep cluster surveys.
\end{abstract}

\section{Deep X--ray Cluster Surveys and the BMW Project}
In the last few years, X--ray selected samples of clusters of galaxies have 
become a formidable tool for cosmology.  Deep surveys using ROSAT PSPC
archival data 
have been used to study the evolution of the cluster abundance and X--ray 
luminosity function (XLF, e.g. Borgani et al. 1999).  
The lack of evolution observed for $L\lsim L^*\simeq 4\cdot 10^{44}\,
h^{2}$ erg s$^{-1}$ 
out to $z\sim 0.8$ favours low values for $\Omega_M$ under reasonable
assumptions about the evolution of the $L-T$ relation.  At the same
time, the original evidence from the EMSS (Henry et al. 1992) of
significant evolution at the very bright end of the XLF has been
confirmed (Vikhlinin  et al. 1998, Rosati et al. 2000 and references
therein).  The main statistical limitation of this conclusion rests with
the small sky coverage of the ROSAT deep surveys, which
clashes with the intrinsic rarity of highly luminous clusters.
XMM--Newton and Chandra are already attracting justified 
attention as the  
likely source for future samples, but no significant sets of serendipitously 
selected clusters can be reasonably expected from these observatories
for at least another 2 
years.  This presents the window of opportunity for our survey, which uses
data from the ROSAT High--Resolution Imager (HRI) archive. With respect 
to the PSPC, the HRI has lower sensitivity and higher instrument background.
However, the HRI offers superior angular resolution, and the archive contains 
many very long observations.  Our results indicate that it is
actually a surprisingly good source of samples of high-redshift
clusters.

\begin{figure}
\centering
\includegraphics[width=.6\textwidth]{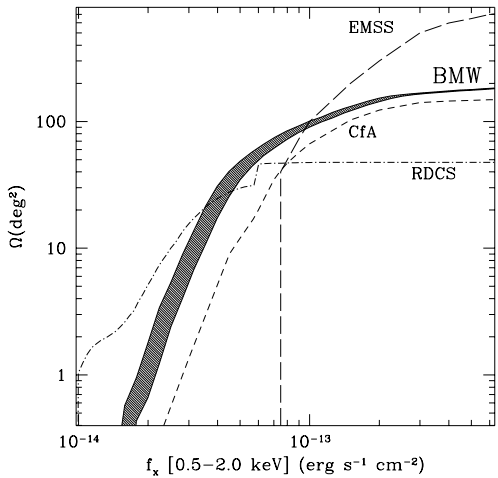}
\caption[]{Sky coverage of the BMW fields for typical extended
sources, compared to some previous X--ray cluster surveys (see Rosati
et al. 2000 for relevant references).  Note the good compromise
reached between a fairly 
large area at intermediate fluxes ($\sim 100$ sq. deg. around 
$\sim 10^{-13}$ erg cm$^{-2}$ s$^{-1}$), and the depth of the sample
(1 sq. deg. at $2.5 \times 10^{-14}$ erg cm$^{-2}$ s$^{-1}$).  The
overall sky coverage is not only a function of flux, but depends also
on the source extension.  Being constructed from a different
instrument with greater angular resolution, the BMW sample provides an
important independent check on possible biases inherent in the PSPC
data used for most surveys so far} 
\label{skycov}
\end{figure}

The Brera Multi-scale Wavelet (BMW) project has currently completed the
systematic 
analysis of about 3100 HRI pointings using a multi--scale wavelet algorithm
(Lazzati et al. 1999). This resulted in a catalog of $\sim 19000$
serendipitous sources with measured fluxes and extensions (Campana et al. 1999, Panzera et al. 2000). 
A clever selection of the HRI energy channels produced a reduction
of the background noise by a factor of $\sim 3$, thus greatly
improving the ability to detect low--surface--brightness sources as clusters.
Cluster candidates were isolated on the basis of their extension,
selecting at a high significance level (corresponding to $>5 \sigma$)
and using only the well--sampled HRI area between 3 and 
15 arcmin off-axis.  After excluding fields with $|b_{II}| \le 20$ or pointed 
on the LMC and SMC, we ended up with a list of more than 500
candidates, which were visually classified on the DSS2 to reject obvious 
contaminants (95 rejections, mostly multiple detections of
substructures in nearby, very extended Abell clusters, plus a few
nearby galaxies).   Our present goal is to complete first the identification
of a high--quality sample of 164 candidates that remain after
excluding cluster--targeted HRI 
fields (to avoid the bias produced by the cluster--cluster angular 
correlation function, for which we have a clear positive detection in
these fields) and then selecting the deeper HRI exposures ($t\ge$ 
10,000 sec).   80\% of the sources in this sample show no counterpart
in the DSS2 and are therefore strong candidates for being distant ($z>0.4$) 
clusters.   Optical follow--up is now underway with telescopes in
both emispheres (mostly the TNG in La Palma and the ESO 3.6~m
telescopes), using multi-band imaging to detect the presence of a
galaxy overdensity while confirming its reality through photometric
redshifts.   We have recently reached a total of
almost 40 candidates for which deep imaging has been secured.
Preliminary analysis of these observations suggests a success rate
(i.e. evidence for a galaxy overdensity correlated with the X--ray
source) of about 80\%.  For the still unidentified 20\% fraction, we
plan further NIR imaging using SOFI at NTT and SQIID at the 4~m KPNO
telescope. 

In parallel, the very last $\sim 1000$ HRI fields from the archive (several with long exposures), are currently being 
analysed and are expected to contribute an extra $\sim 40\%$ average
increase in the sample size and sky coverage.

\begin{figure}
\centering
\includegraphics[width=.6\textwidth]{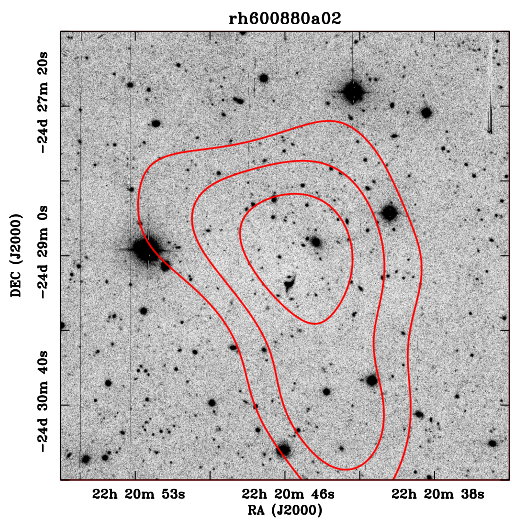}
\caption[]{$R$ image (15~min, ESO 3.6~m + EFOSC2) of BMW222045-242, a
cD--dominated poor cluster at $z\simeq 0.5$ identified during the first
follow--up run of the BMW candidates
} 
\label{cluster}
\end{figure}

\clearpage
\addcontentsline{toc}{section}{Index}
\flushbottom
\printindex


\begin{thebibliography}{7}
%
\addcontentsline{toc}{section}{References}
\bibitem{} Borgani, S., et al., 1999, ApJ, 517, 40.
\bibitem{} Campana, S., et al., 1999, ApJ,
524, 423.
\bibitem{} Henry, P. et al., 1992, ApJ, 386, 408.
\bibitem{} Gioia, I.M., et al. 1990, ApJS, 72, 567.
\bibitem{} Lazzati, D., et al., 1999, ApJ, 524, 414.
\bibitem{} Panzera, M.R., et al., 2000, in preparation
\bibitem{} Rosati, P., et al. 1998, ApJ, 492, L21.
\bibitem{} Rosati, P., et al. 2000, in ``Large-scale structure in the X-ray
Universe'', in press (astro-ph/0001119).
\bibitem{} Vikhlinin, A., et al. 1998a, ApJ, 498, L21.
\bibitem{} Vikhlinin, A., et al. 1998b, ApJ, 502, 558.

\end{thebibliography}
\end{document}